\documentclass[english,12pt]{article}
          \usepackage{babel}
          \usepackage[T1]{fontenc}
          \usepackage[latin2]{inputenc}
          \usepackage{pslatex}
\begin{document}

\title{\bf Curious Variables Experiment (CURVE). \\
IX Draconis - a Clue for Understanding Evolution of Cataclysmic
Variable Stars}
\author{A. ~O~l~e~c~h$^1$, ~K. ~Z~{\l}~o~c~z~e~w~s~k~i$^2$, 
~K. ~M~u~l~a~r~c~z~y~k$^2$, ~P. ~K~\c{e}~d~z~i~e~r~s~k~i$^2$,\\
~M. ~W~i~{\'s}~n~i~e~w~s~k~i$^1$ ~and~ ~G. ~S~t~a~c~h~o~w~s~k~i$^1$}
\date{$^1$ Nicolaus Copernicus Astronomical Center,
Polish Academy of Sciences,
ul.~Bartycka~18, 00-716~Warszawa, Poland\\
{\tt e-mail: (olech,mwisniew,gss)@camk.edu.pl}\\
~\\
$^2$ Warsaw University Observatory, Al. Ujazdowskie 4, 00-476 Warszawa,
Poland\\ {\tt e-mail: (kzlocz,kmularcz,pkedzier)@astrouw.edu.pl}}
\maketitle

\begin{abstract} 

We report extensive photometry of frequently outbursting dwarf nova IX
Draconis. During five months of observations the star went into three
superoutbursts and seven ordinary outbursts. This allowed us to
determine its supercycle and cycle lengths as equal to $54\pm1$ and
$3.1\pm0.1$ days, respectively. During the September 2003 superoutburst,
which had the best observational coverage, IX Dra displayed clear
superhumps with a period of $P_{sh}=0.066968(17)$ days ($96.43\pm0.02$
min). This period was constant during the whole superoutburst. Another
period, which was clearly present in the light curve of IX Dra in
superoutburst, had a value of 0.06646(6) days ($95.70\pm0.09$ min) and
we interpret it as the orbital period of the binary. Thus IX Dra is the
first SU UMa star showing orbital modulation during the entire
superoutburst. The beat between these two periods is the main cause of an
unusual phase reversal of superhumps - a phenomenon which was
previously observed in ER UMa. If our interpretation of the second
periodicity is correct, IX Dra has an extremely low period excess
$\epsilon$ equal to only $0.76\% \pm 0.03$\%. This implies very low mass
ratio $q=0.035\pm0.003$, which strongly suggests that the system
contains a brown dwarf-like degenerate secondary of mass $\sim0.03 ~\cal
M_\odot$ and that IX Dra is the most evolved dwarf nova known.

Such a very low mass ratio results in the outer edge of the accretion
disk reaching 80\% of the distance between the components of the system.
In turn, this allows the disk particles to enter a 2:1 resonance and
leads to the appearance of the orbital period in the light curve of the
entire superoutburst.

The high level of activity and brightness of IX Dra indicate that very
old cataclysmic variables go through episodes of increased activity
leading to loss of angular momentum through mass loss from the system.

Modulations with the orbital period are also detectable during normal
outbursts and in quiescence.

\end{abstract}

\section{Introduction}

Dwarf novae are non-magnetic cataclysmic variables, which are close
binary systems containing white dwarf primary and Roche lobe filling
secondary. The secondary is typically a low mass main sequence star,
which loses its material through the inner Lagrangian point. This
material forms an accretion disc around a white dwarf (Warner 1995).

One of the most intriguing classes of dwarf novae are SU UMa stars which
have short orbital periods (less than 2.5 hours) and show two types of
outbursts: normal outbursts and superoutbursts. Superoutbursts are
typically about one magnitude brighter than normal outbursts, occur
about ten times less frequently and display characteristic tooth-shape
light modulations with a period a few percent longer than the orbital
period of the binary.

The behavior of SU UMa stars in now well understood within the frame of
the thermal-tidal instability model (see Osaki 1996 for review). Superhumps
occur at a period slightly longer than the orbital period of the binary
system. They are most probably the result of accretion disc precession
caused by gravitational perturbations from the secondary. These
perturbations are most effective when disc particles moving in
eccentric orbits enter the 3:1 resonance. Then the superhump period is
simply the beat period between orbital and precession rate periods.

Over ten years ago, the diversity of behavior observed in cataclysmic
variables seemed to be quite small. These variables were classified into
few distinct groups which were clearly visible in the $P_{orb}-\dot M$
diagram (see for example Fig. 3 of Osaki 1996). There is a group of SU
UMa stars located below the period gap. At longer orbital periods one
can find three other groups: U Gem, Z Cam and nova-like stars, listed
according to increasing $\dot M$.

At the beginning of the 1990s, the situation started to be more complicated. 
The systems showing superhumps were divided into four subgroups: 

\begin{itemize} 

\item ordinary SU UMa stars,

\item permanent superhumpers - high accretion rate systems permanently in
superoutbursts (Skillman and Patterson 1993),

\item WZ Sge stars, characterized by an extremely long quiescent state,
going into superoutburst every $\sim$10 years and showing no or infrequent
ordinary outbursts,

\item ER UMa stars - systems characterized by an extremely short
supercycle (20-60 days), a short interval between normal outbursts
(3-4 days) and small amplitude (3 mag) of superoutbursts (Kato and
Kunjaya 1995, Robertson et al. 1995)

\end{itemize}

There are only five known ER UMa stars: ER UMa itself (Kato and Kunjaya
1995, Robertson et al. 1995), V1159 Ori (Robertson et al. 1995,
Patterson et al. 1995), RZ LMi (Robertson et al. 1995, Nogami et al.
1995), DI UMa (Kato et al. 1996) and IX Dra (Ishioka et al. 2001). The
most poorly observed object in this group is IX Dra, thus we decided
to include it in the list of stars observed within the Curious Variables
Experiment (Olech et al. 2003a,b). Here we report the results of a five
month observational campaign performed in 2003.

\section{IX Draconis}

The variability of IX Dra was discovered by Noguchi et al. (1982). The
light curve obtained by Klose (1995) allowed him to determine the
amplitude of light variations as equal to 3.5 mag and its period as
equal to 45.7 days. These properties indicated that the star could be a
CV system and thus IX Dra was classified as U Gem-type variable in the
CV catalog of Downes et al. (1997). This classification was confirmed by
spectroscopic observations of Liu et al. (1999).

IX Dra was extensively observed in 2000-2001 by Ishioka et al. (2001).
Their observations revealed that the star is a member of the ER UMa-type
group of variables, having a supercycle length of 53 days; an interval
of normal outbursts of 3-4 days; a duty cycle $\sim30$\% and an outburst
amplitude of 2.5 mag. From time-series observations during the
superoutbursts, they obtained a presumable superhump period of 0.067
days.

\section{Observations and Data Reduction}

Observations of IX Dra reported in the present paper were obtained during
46 nights between June 24, 2003 and November 22, 2003 at the
Ostrowik station of the Warsaw University Observatory. The data were
collected using the 60-cm Cassegrain telescope equipped with a
Tektronics TK512CB back-illuminated CCD camera. The scale of the camera
was 0.76"/pixel providing a 6.5'$\times$6.5' field of view. The full
description of the telescope and camera was given by Udalski and Pych
(1992).

\vspace{10.8cm}

\includegraphics{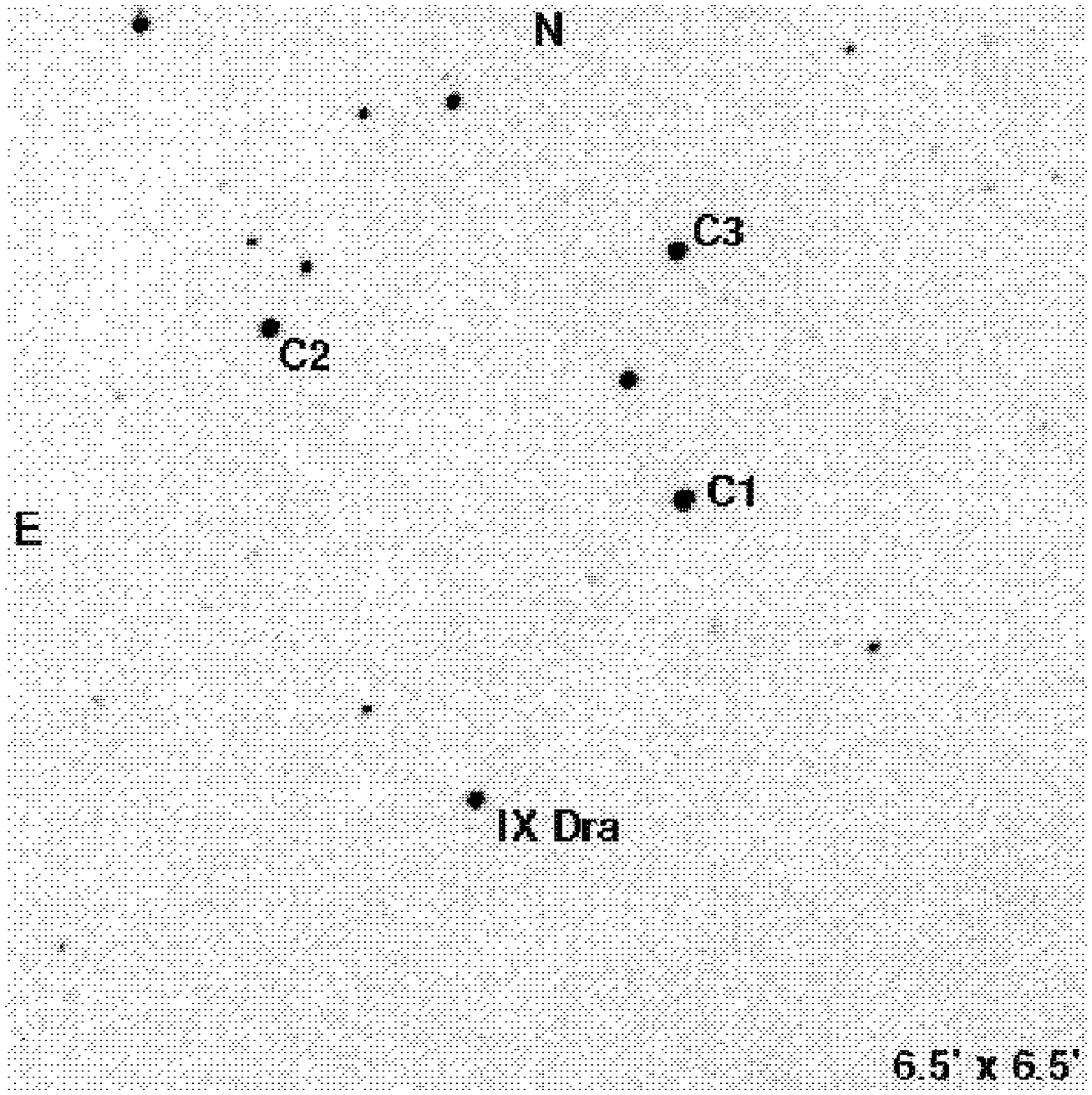}

   \begin{figure}[h]
      \caption{\sf Finding chart for IX Dra covering a region of $6.5 \times
6.5$ arcminutes. The positions of the comparison stars are shown. North
is up, East to the left.}
   \end{figure}

We monitored the star in ``white light'' in order to be able to
observe it also at minimum light of around 17.5 mag.

The exposure times were from 90 to 180 seconds during the bright state
and from 150 to 350 seconds at minimum light.

A full journal of our CCD observations of IX Dra is given in Table 1. In
total, we monitored the star for 141 hours and obtained 2759
exposures.

\begin{table}[!h]
\caption{\sc Journal of the CCD observations of IX Dra.}
\vspace{0.1cm}
\begin{center}
{\tiny
\begin{tabular}{|l|c|c|r|r|}
\hline
\hline
Date of& Start & End & Length & No. of \\
2003   & 2452000. + & 2452000. + & [hr]~ & frames \\
\hline
Jun 24/25 &815.35902 &815.37233 &0.319 &5\\
Jun 25/26 &816.42810 &816.43371 &0.135 &4\\
Jun 26/27 &817.44616 &817.51757 &1.714 &43\\
Jun 28/29 &819.39763 &819.46972 &1.730 &37\\
Jun 30/01 &821.41308 &821.51691 &2.492 &32\\
Aug 04/05 &856.46990 &856.56842 &2.364 &60\\
Aug 05/06 &857.50050 &857.56902 &1.644 &46\\
Aug 06/07 &858.31611 &858.57326 &3.677 &101\\
Aug 08/09 &860.32764 &860.49241 &3.954 &60\\
Aug 16/17 &868.37691 &868.53273 &3.740 &96\\
Aug 17/18 &869.36583 &869.45143 &2.054 &47\\
Aug 20/21 &872.39664 &872.48818 &2.197 &41\\
Aug 22/23 &874.37718 &874.51764 &3.371 &61\\
Aug 23/24 &875.53634 &875.59433 &1.392 &34\\
Aug 24/25 &876.55043 &876.55399 &0.085 &3\\
Aug 25/26 &877.40600 &877.51050 &2.508 &38\\
Aug 27/28 &879.46050 &879.56141 &2.421 &47\\
Aug 28/29 &880.31593 &880.36075 &1.076 &23\\
Aug 30/31 &882.31985 &882.50416 &4.423 &124\\
Aug 31/01 &883.33636 &883.49807 &3.881 &62\\
Sep 01/02 &884.47388 &884.55306 &1.900 &10\\
Sep 02/03 &885.51654 &885.60859 &2.209 &47\\
Sep 03/04 &886.32950 &886.54790 &5.242 &111\\
Sep 05/06 &888.40253 &888.55522 &3.665 &93\\
Sep 06/07 &889.30712 &889.56523 &6.195 &141\\
Sep 07/08 &890.31662 &890.56445 &5.948 &133\\
Sep 13/14 &896.30359 &896.54744 &5.852 &126\\
Sep 14/15 &897.29154 &897.61492 &2.811 &76\\
Sep 15/16 &898.56906 &898.59302 &0.575 &17\\
Sep 16/17 &899.50051 &899.51023 &0.233 &5\\
Sep 20/21 &903.43127 &903.60900 &4.265 &76\\
Sep 21/22 &904.24650 &904.60668 &8.644 &161\\
Sep 22/23 &905.23879 &905.58645 &8.344 &151\\
Sep 24/25 &907.24152 &907.58795 &8.314 &127\\
Sep 25/26 &908.24563 &908.63513 &9.348 &172\\
Sep 26/27 &909.24138 &909.63448 &9.434 &154\\
Sep 30/01 &913.35124 &913.51222 &0.681 &13\\
Oct 01/02 &914.43357 &914.48948 &1.342 &24\\
Oct 03/04 &916.24779 &916.30463 &1.364 &20\\
Oct 15/16 &928.32350 &928.34468 &0.508 &12\\
Oct 18/19 &931.26054 &931.37863 &1.146 &24\\
Oct 19/20 &932.32327 &932.37064 &1.137 &14\\
Oct 31/01 &944.40016 &944.41475 &0.350 &6\\
Nov 19/20 &963.63478 &963.66356 &0.691 &9\\
Nov 20/21 &964.34627 &964.66717 &2.042 &15\\
Nov 22/23 &966.20520 &966.70929 &3.643 &58\\
\hline
Total          &   --   & -- & 140.1 & 2759 \\ 
\hline
\hline
\end{tabular}}
\end{center}
\end{table}

All the data reductions were performed using a standard procedure
based on the IRAF\footnote{IRAF is distributed by the National Optical
Astronomy Observatory, which is operated by the Association of
Universities for Research in Astronomy, Inc., under a cooperative
agreement with the National Science Foundation.} package and
profile photometry was derived using the DAOphotII package
(Stetson 1987).

Relative unfiltered magnitudes of IX Dra were determined as the
difference between the magnitude of the variable and the intensity
averaged magnitude of three comparison stars shown on finding chart in
Fig. 1. The magnitudes and colors of our comparison stars were taken
from Henden and Honeycutt (1995) and are summarized in Table 2.

\begin{table}[!h]
\caption{\sc Properties of comparison stars used for calibration of IX Dra
magnitude.}
\begin{center}
\begin{tabular}{|l|c|c|c|c|}
\hline
\hline
Star & R.A. & Decl. & $V$ & $B-V$ \\
     & 2000.0 & 2000.0 & & \\
\hline
C1 & $18^h12^m19.6^s$ & $67^\circ 06'26.7"$ & 14.524 & 1.116\\
C2 & $18^h12^m43.4^s$ & $67^\circ 07'24.6"$ & 14.652 & 0.704\\
C3 & $18^h12^m19.9^s$ & $67^\circ 07'20.1"$ & 14.727 & 0.463\\
\hline
\hline
\end{tabular}
\end{center}
\end{table}

The typical accuracy of our measurements varied between 0.004 and 0.090
mag depending on the brightness of the object. The median value of the
photometric errors was 0.015 mag.

The $B-V$ color of outbursting novae is $\sim0$ mag (Bruch and Engel 1994).
Frequently outbursting ER UMa stars have $B-V \approx 0$ even in
quiescence due to large mass transfer rates. Thus the most appropriate
star for transforming the relative light curve to the Johnson $V$ system
is the bluest star, $C3$. Our observations were made in
"white light", which roughly corresponds to Cousins $R$ band (Udalski
and Pych 1992). Knowing the $B-V$ color of $C3$ we can estimate its
$V-R$ color which is $\sim0.27$ mag (Caldwell et al. 1993). Using this
value and $V$ magnitude of $C3$ we transformed our relative light curve
to the Johnson $V$ system.

\section{General light curve}

The nightly light curves were intensity averaged into one-day bins for
runs shorter than four hours and in half-day bins for longer runs. The
resulting light curve for the period August-November 2003 is shown in Fig.
2. In this interval, the star displayed three superoutbursts and six
ordinary outbursts. A seventh outburst, not shown in Fig. 2, was
additionally observed at the end of June 2003.

\vspace{8.6cm}

\includegraphics{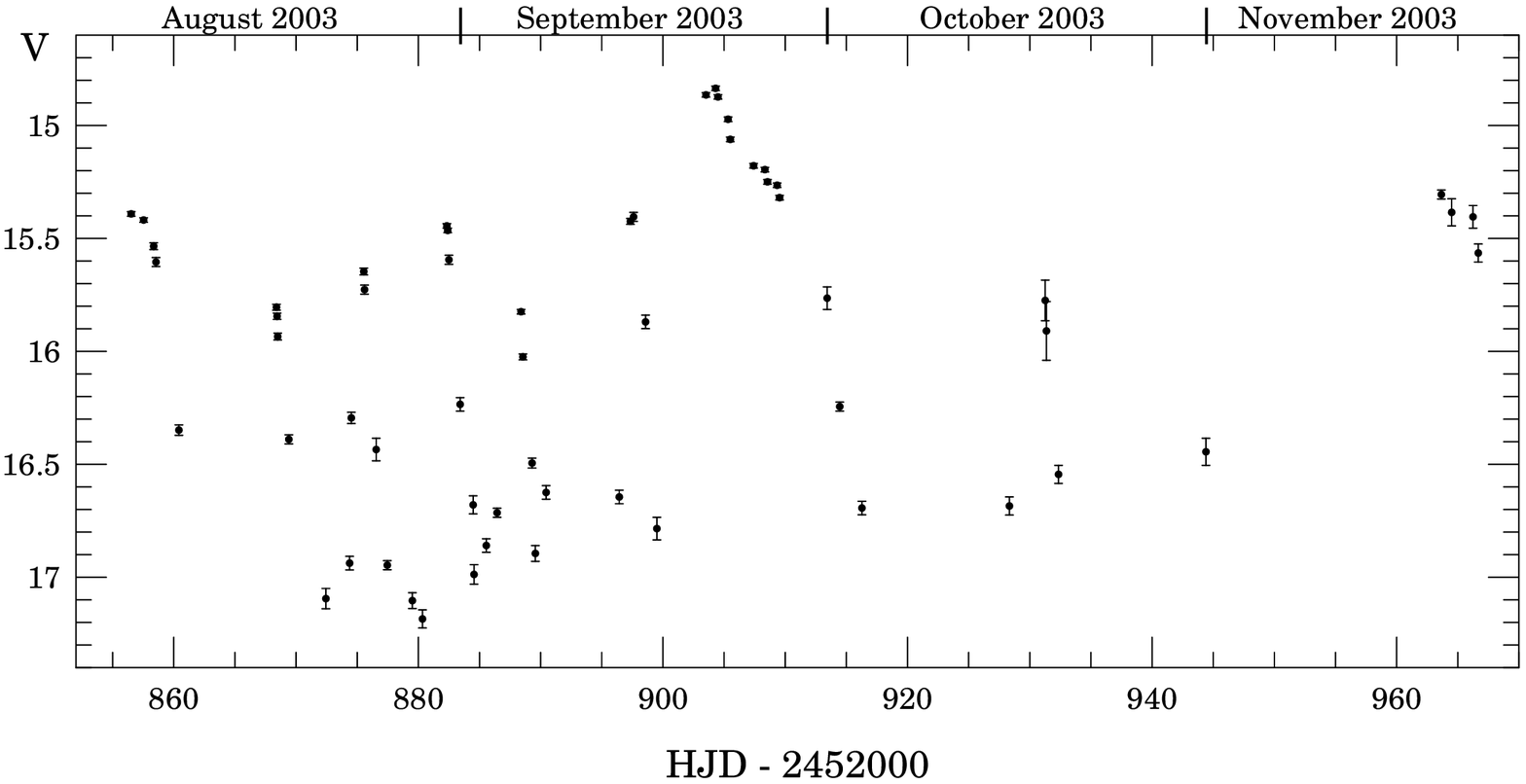}

   \begin{figure}[h]
      \caption{\sf The general photometric behavior of IX Dra during
our 2003 campaign.}
   \end{figure}

The minimal brightness of the star in quiescence is 17.3 mag. In the
superoutburst, IX Dra reaches 14.8 mag, and in ordinary outbursts around
15.4 mag.

The general light curve of IX Dra was analyzed using {\sc anova}
statistics with two harmonic Fourier series (Schwarzenberg-Czerny 1996).
The resulting periodogram, for the frequency range $0\div0.6$ c/d, is shown
in Fig. 3. Two dominant peaks correspond to the periods $54\pm1$ and
$3.1\pm0.1$ days, which can be interpreted as intervals between two
consecutive superoutbursts and ordinary outbursts, respectively. These
values are in very good agreement with results obtained by Ishioka et
al. (2001). One can note the increasing length of the supercycle: in the first
half of the 1990s it was 45.7 days (Kolb 1995), already
53 days (Ishioka et al. 2001) in the years 2000-2001 and 54 days in 2003. 
Changes of the supercycle length were observed in other ER UMa-type variables. 
ER UMa itself showed increase of supercycle with rate of $\dot P \approx
4\times 10^{-3}$ and RZ LMi decrease with rate of $\dot P=-1.7\times
10^{-3}$ (Robertson et al. 1995).

\vspace{8.8cm}

\includegraphics{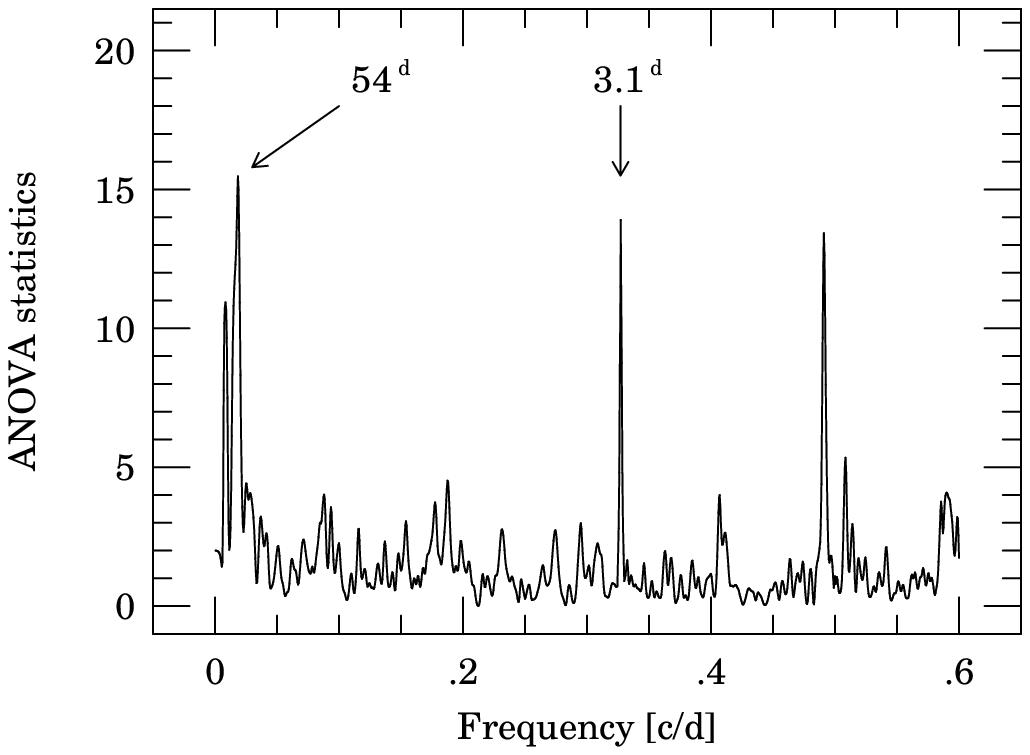}

   \begin{figure}[h]
      \caption{\sf The power spectrum of IX Dra global light curve. The
arrows mark the position of most prominent peaks corresponding to
periods 54 and 3.1 days.
              }
   \end{figure}

The general light curve of IX Dra in superoutbursts was folded on a
supercycle period of 54 days and is displayed in Fig. 4. It is clear
that the whole superoutburst lasts around 13 days and is divided into a
quick initial rise ($<1$ day), a plateau phase ($\sim 10$ days) and a final
decline ($\sim 3$ days). The decline rate is $0.078(2)~{\rm mag\cdot
d}^{-1}$ and $0.26(2)~{\rm mag\cdot d}^{-1}$ during plateau and final
decline phases, respectively.

The overall behavior of IX Dra is very consistent with the model of ER
UMa-type variables computed by Osaki (1995). His bolometric light curve
(see his Fig. 2), obtained for binary system with mass transfer rate of
$\dot M=4.0\times 10^{16}~{\rm g\cdot s}^{-1}$, resembles our
observations in all details. The model light curve shows superoutbursts 
occurring every 44 days, the duration of a superoutburst of 20 days and
the  recurrence time of normal outbursts of 4-5 days.

\clearpage

~
\vspace{5.5cm}

\includegraphics{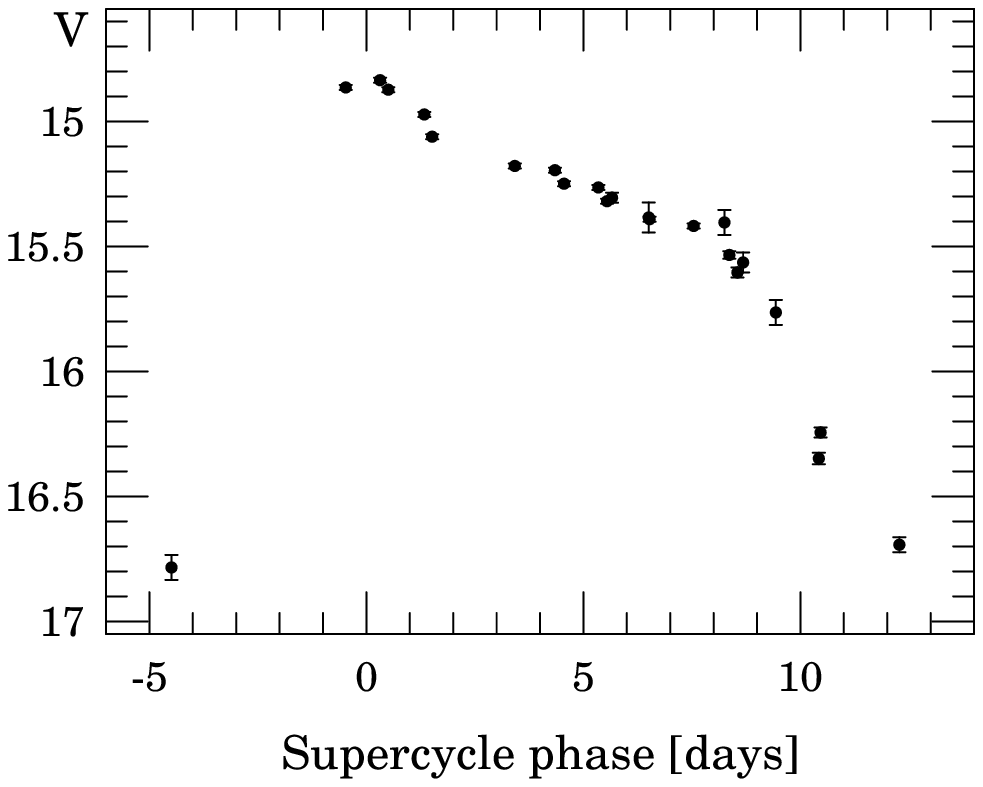}

   \begin{figure}[h]
      \caption{\sf The light curve of IX Dra in superoutburst obtained
by folding the general light curve with supercycle period of 54 days.
              }
   \end{figure}

\vspace{13.4cm}

\includegraphics{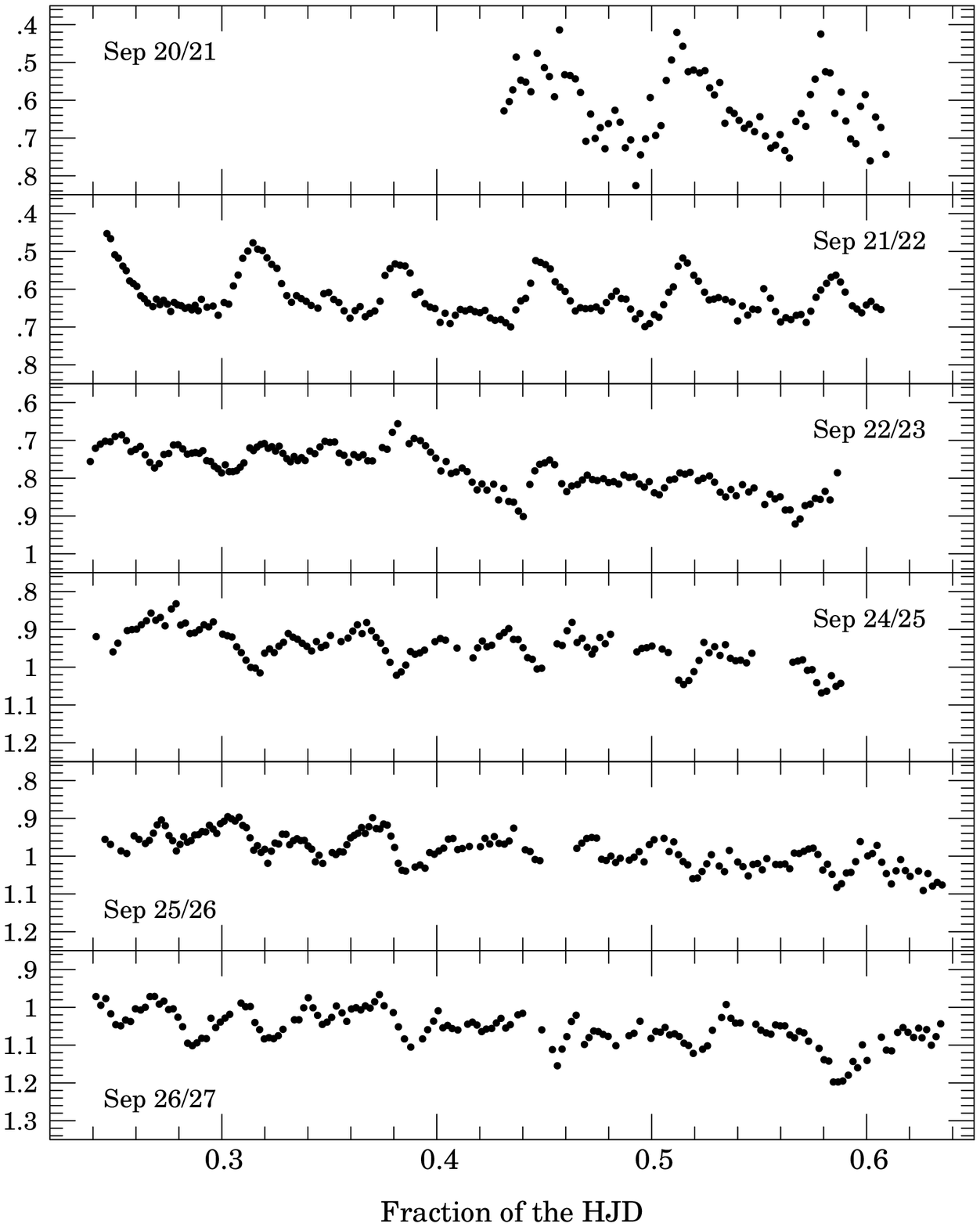}

   \begin{figure}[h]
      \caption{\sf The light curves of IX Dra during its 2003 September
superoutburst.
              }
   \end{figure}

\clearpage

\section{The September 2003 superoutburst}

Our observations from Sep 20/21 caught IX Dra in a very bright state. Three
nights earlier the star was in quiescence. The mean brightness of IX Dra
at the beginning of the night of Sep 21/22 was even higher than on Sep
20/21, thus we conclude that the superoutburst started on Sep 20.

Light curve from nights Sep 20/21 and 21/22 shows well developed
superhumps with an amplitude of 0.27 and 0.16 mag, respectively (see
Fig. 5). On Sep 21, the clear secondary humps became visible. One night
later, the amplitude of superhumps declined to 0.08 mag, and secondary
humps were even more prominent, having an amplitude only slightly
smaller than the main maxima. During the nights Sep 24/25, 25/26 and
26/27 the amplitudes of variability were 0.09, 0.08 and 0.08 mag,
respectively. The shape of the superhump profile became complex and it
was very difficult to say which maximum was the primary and which the
secondary.

\subsection{The $O-C$ analysis}

To check the stability of the superhump period and to determine its
value we constructed an $O-C$ diagram. We decided to use the timings of
primary minima, because they were almost always deep and clearly visible
in the light curve of the variable. In the end, we were able to
determine 29 times of primary minima and they are listed in Table 3
together with their errors, cycle numbers $E$ and $O-C$ values.

\begin{table}[!h]
\caption{\sc Times of minima in the light curve of IX Dra during its
2003 September superoutburst.}
\vspace{0.1cm}
\begin{center}
\begin{tabular}{|c|c|c|r|}
\hline
\hline
Cycle & $HJD_{\rm min}-2452000$ & Error & $O-C$\\
number $E$ & & & [cyles] \\
\hline
0 & 903.4940 & 0.0045 &  $-0.0388$\\
1 & 903.5635 & 0.0030 &  $-0.0009$\\
12 & 904.3000 & 0.0030 & $-0.0023$\\
13 & 904.3670 & 0.0040 & $-0.0018$\\
14 & 904.4340 & 0.0025 & $-0.0012$\\
15 & 904.4990 & 0.0030 & $-0.0305$\\
16 & 904.5700 & 0.0040 & $ 0.0297$\\
27 & 905.3045 & 0.0025 & $-0.0015$\\
28 & 905.3710 & 0.0035 & $-0.0084$\\
29 & 905.4395 & 0.0025 & $ 0.0145$\\
30 & 905.5030 & 0.0040 & $-0.0372$\\
31 & 905.5682 & 0.0040 & $-0.0635$\\
57 & 907.3177 & 0.0030 & $ 0.0629$\\
58 & 907.3822 & 0.0030 & $ 0.0261$\\
59 & 907.4490 & 0.0040 & $ 0.0236$\\
60 & 907.5150 & 0.0025 & $ 0.0093$\\
61 & 907.5806 & 0.0030 & $-0.0111$\\
71 & 908.2550 & 0.0060 & $ 0.0601$\\
72 & 908.3205 & 0.0040 & $ 0.0383$\\
73 & 908.3870 & 0.0030 & $ 0.0314$\\
74 & 908.4560 & 0.0050 & $ 0.0618$\\
75 & 908.5200 & 0.0030 & $ 0.0176$\\
76 & 908.5870 & 0.0025 & $ 0.0181$\\
86 & 909.2520 & 0.0025 & $-0.0510$\\
87 & 909.3220 & 0.0025 & $-0.0057$\\
88 & 909.3900 & 0.0030 & $ 0.0098$\\
89 & 909.4560 & 0.0020 & $-0.0046$\\
90 & 909.5210 & 0.0030 & $-0.0339$\\
91 & 909.5870 & 0.0030 & $-0.0483$\\
\hline
\hline
\end{tabular}
\end{center}
\end{table}

The least squares linear fit to the data from Table 3 gives the
following ephemeris for the minima:

\begin{equation}
{\rm HJD}_{min} =  2452903.3966(12) + 0.066963(19) \cdot E
\end{equation}

The $O-C$ values computed according to the ephemeris (1) are listed in
Table 3 and also shown in Fig. 6. It is clear that there is no trace of
period change, thus we conclude that period of the superhumps during
2003 September superoutburst of IX Dra was constant and its value was
$P_{sh}=0.066963(19)$ days. 

\vspace{7.2cm}

\includegraphics{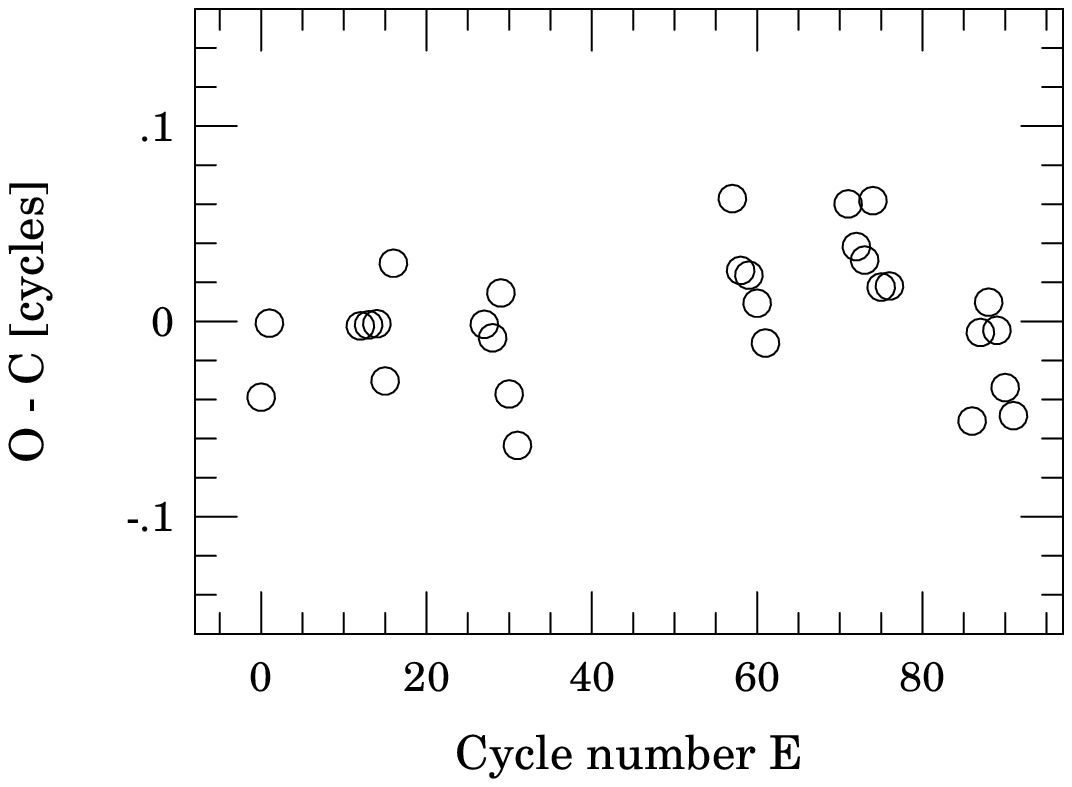}

   \begin{figure}[h]
      \caption{\sf The $O-C$ diagram for superhumps minima of IX Dra
detected during its 2003 September superoutburst.
              }
   \end{figure}

\subsection{Nightly light curves}

Knowing the period of superhumps we can phase the nightly light curves
from superoutburst to trace the superhump profile changes. The result of
such an operation is shown in Fig. 7. Phase 0.0 corresponds to a deep
minimum in the light curve. It is clear that at the beginning of
superoutburst, the star displayed large-amplitude tooth-shape superhumps
with no secondary maxima. One night later, the amplitude significantly
decreased, and secondary maxima became visible. On Sep 22, the amplitude
of secondary humps increased, and on Sep 24 they became higher than the
main maxima. The situation repeated on nights Sep 25 and Sep 26.

\vspace{14cm}

\includegraphics{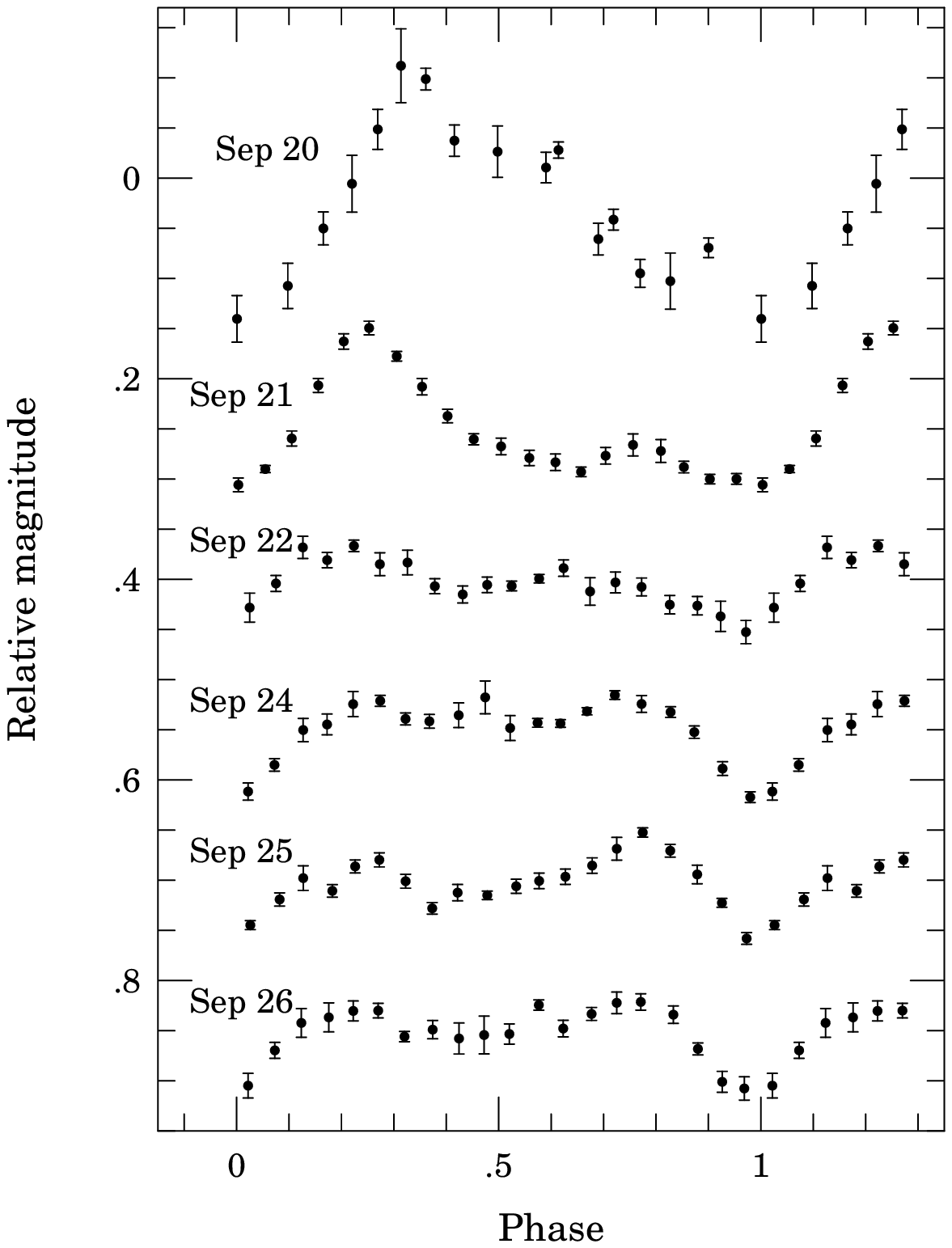}

   \begin{figure}[h]
   \centering
      \caption{\sf Amplitude and profile changes of superhumps during 2003
September superoutburst of IX Dra.
              }
   \end{figure}

Such an unexpected phase reversal of maxima during the very early stage
(only five days after the superoutburst maximum) was also observed in
another star from the ER UMa subgroup - ER UMa itself (Kato et al.
2003a). It was interpreted as a sudden switch to so-called late
superhumps usually occurring during late stages of superoutburst in
ordinary SU UMa stars. As we will see later, the origin of this phase
reversal has a completely different source.

\subsection{Additional modulation as a source of phase reversal}

From each light curve of IX Dra in superoutburst we removed the first or
second order polynomial and analyzed them using {\sc anova} statistics
with two harmonic Fourier series (Schwarzenberg-Czerny 1996). The
resulting periodogram is shown in Fig. 8. The most prominent peak is
found at a frequency of $f_1=14.926\pm0.008$ c/d, which corresponds to
the period of $P_{sh}=0.06699(4)$ days. The first harmonic of this
frequency at $f_2=29.86\pm0.05$ c/d is also clearly visible.
\clearpage

~
\vspace{6cm}

\includegraphics{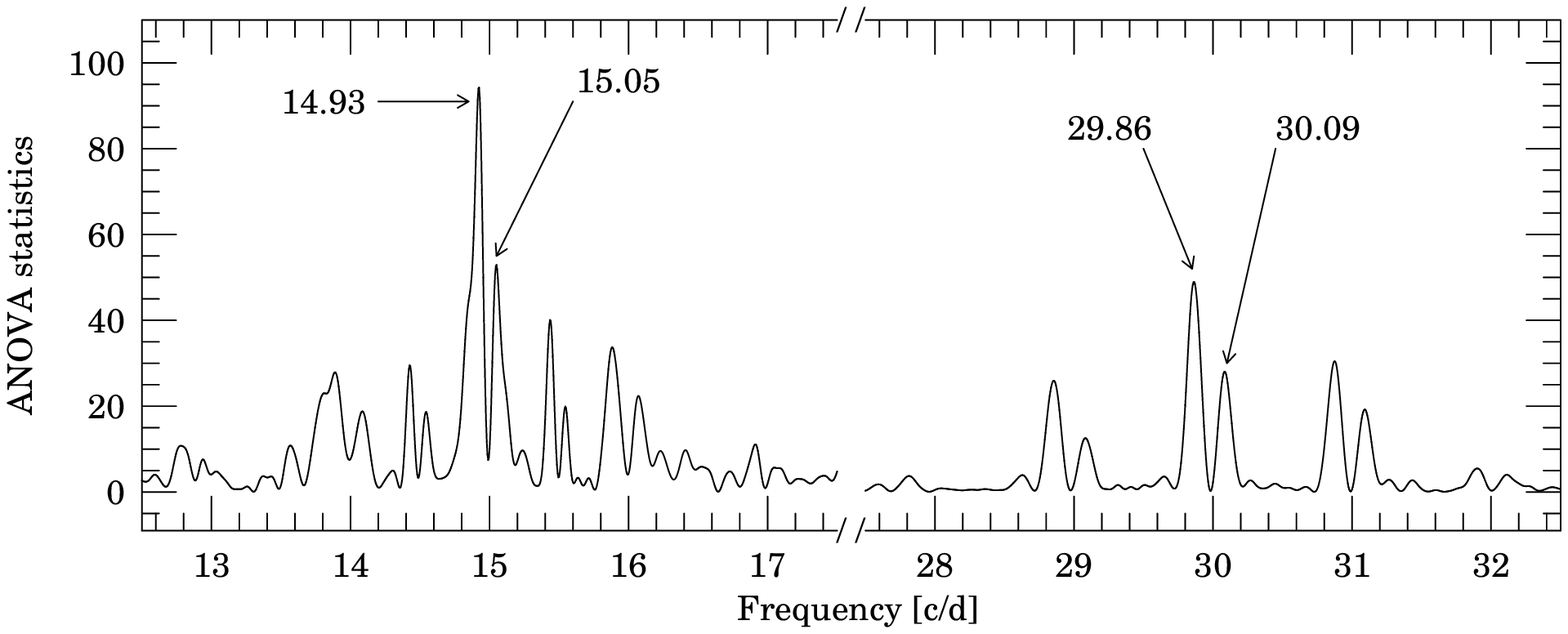}

   \begin{figure}[h]
      \caption{\sf {\sc Anova} power spectrum of the light curve of IX Dra
during its 2003 September superoutburst.
              }
   \end{figure}

Combining this determination with the superhump period obtained from the
$O-C$ analysis, we conclude that the mean superhump period during the
2003 September superoutburst of IX Dra was $P_{sh}=0.066968(17)$ days
($96.43\pm0.02$ min).

A very interesting feature of the power spectrum shown in Fig. 8 is a
double structure of the main peak, which is visible both for the main
frequency and its first harmonics. The secondary peak appears at
$f_3=15.049\pm0.013$ c/d, which corresponds to $P_3=0.06646(6)$ days,
and its first harmonic at $f_4=30.09\pm0.05$ c/d.

The peak at $f_3$ is our {\it Rosetta stone} for understanding the
unusual phase reversal seen in IX Dra and ER UMa. It is easy to find
that the difference between the two main frequencies $f_1$ and $f_3$ is
equal to the beat frequency $f_{beat}=0.123$ c/d. This corresponds to a
period of 8.1 days. This means that at the beginning of superoutbursts
both waves oscillate in one phase, thus we detect clear superhumps with
a large amplitude of 0.27 mag. After half of the beat cycle ($\sim 4$
days) the waves are shifted in phase by 0.5. Thus the maximum of the
modulation characterized by $P_3$ occurs exactly in the place of birth
of the secondary humps. The amplitude of the secondary humps is then
increased by the second wave and they became stronger than the main
maxima.

To check if our interpretation is correct, we decided to perform
prewhitening of the light curve of IX Dra from the superoutburst. The raw
light curve from the period Sep 20-26, with the general decreasing trend
removed, was fitted with two sine series corresponding to the two periods
and their five harmonics:

\begin{equation}
rel.~mag = A_0 + \sum^6_{j=1} A^1_j\sin(2j\pi t/P_{sh}+\phi^1_j) +
\sum^6_{j=1} A^3_j\sin(2j\pi t/P_3+\phi^3_j)
\end{equation}

\noindent Knowing $A^1_j$ and $\phi^1_j$ we were able to remove 
the term containing $P_{sh}$ and obtain pure modulations with $P_3$.
The resulting light curve, folded on the period $P_3=0.06646$ days and
averaged in 0.02 phase bins, is shown in Fig. 9.

On the other hand, knowing $A^3_j$, $\phi^3_j$ and $P_3$ we can remove
the variability with $P_3$ and investigate the pure superhump profiles.
This has been done in Fig. 10, where we show individual light curves from 2003
September superoutburst prewhitened with $P_3$. These light curves are
phased with $P_{sh}$ and averaged in 0.05 phase bins. 

\clearpage

~
\vspace{6cm}

\includegraphics{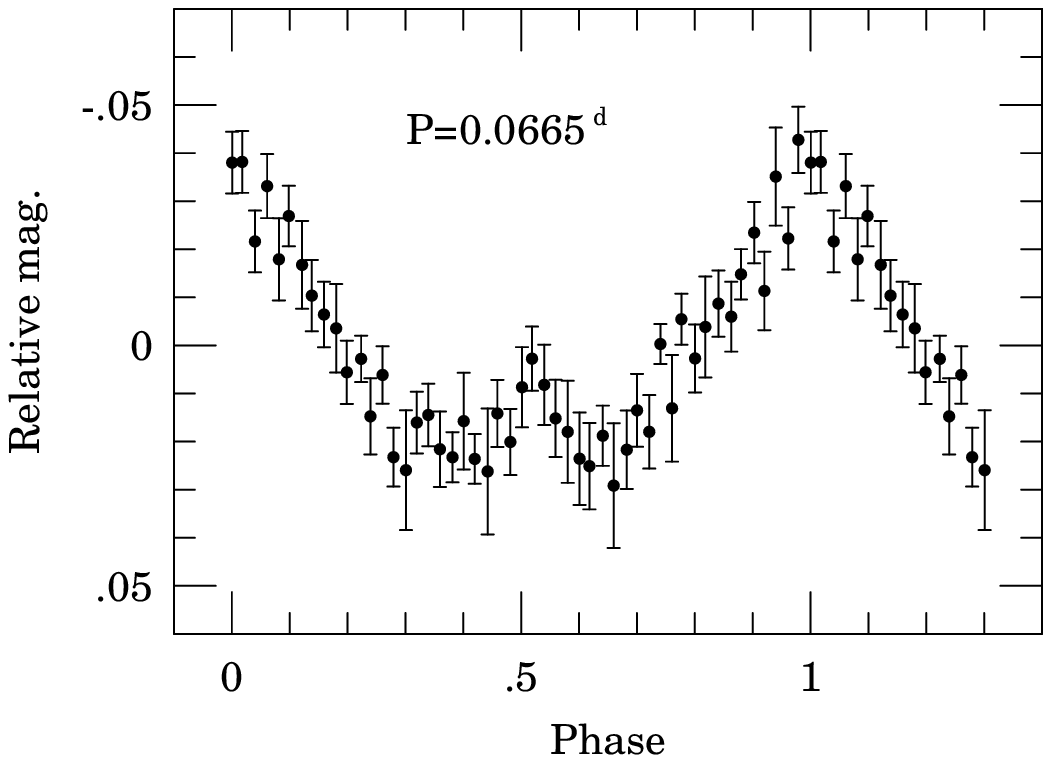}

   \begin{figure}[h]
      \caption{\sf The light curve from 2003 September superoutburst
prewhitenned with $P_{sh}$ and folded with the period of $P_3=0.06646$ days.
              }
   \end{figure}

\vspace{12.6cm}

\includegraphics{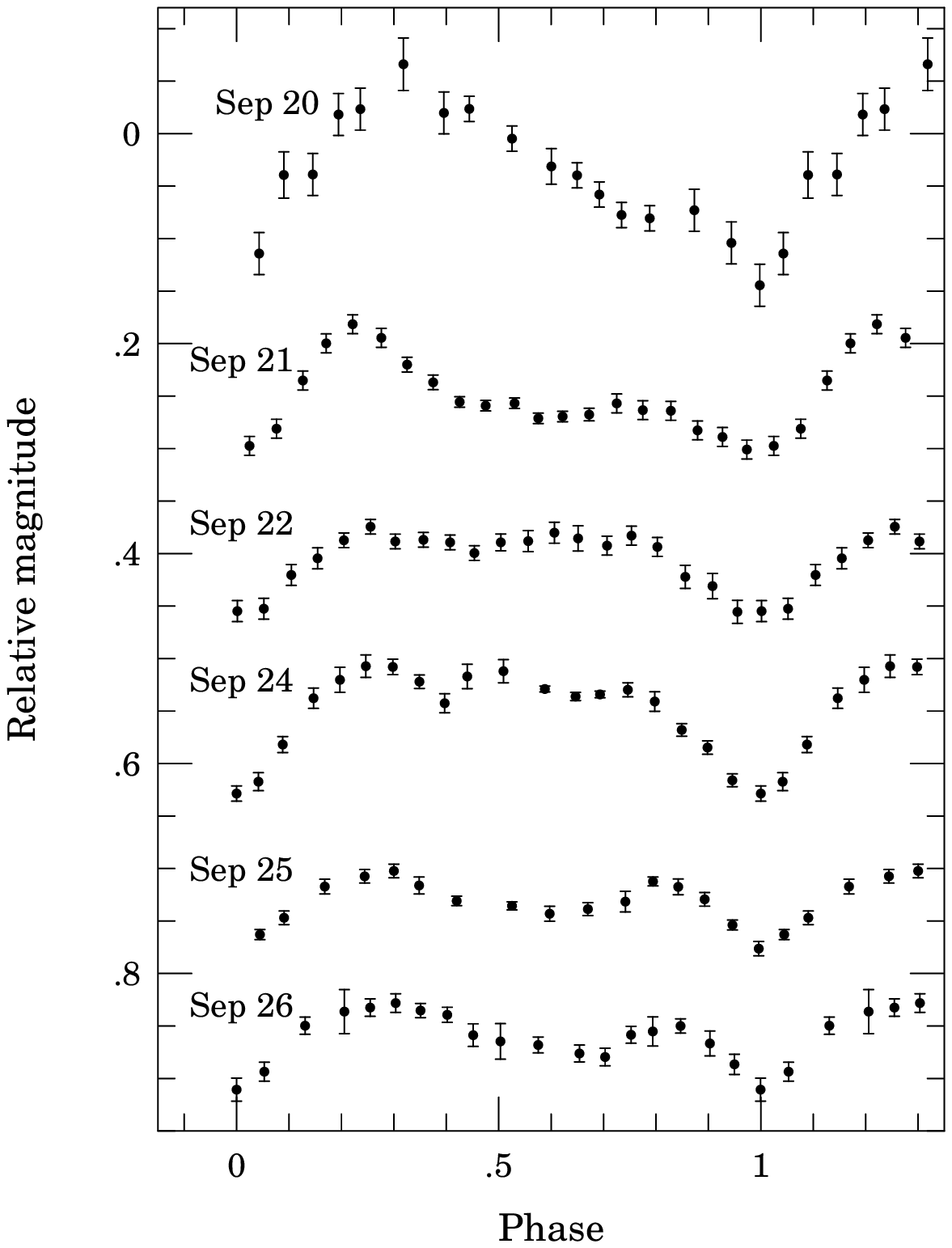}

   \begin{figure}[h]
      \caption{\sf Amplitude and profile changes of superhumps during the 2003
September superoutburst of IX Dra after removing the variability with
the second period present in the light curve ($P_3$).
              }
   \end{figure}

It is now clear that there is no phase reversal, the first maximum
is always the primary one, and superhumps behave as in normal SU UMa star.

The same situation is in ER UMa. Its superhump period, according to Kato
et al. (2003a), is $0.06558(6)$ days ($f=15.25$ c/d). The orbital period
of the system determined from the radial velocity study of Thorstensen et
al. (1997) is $0.06366(3)$ days ($f=15.71$ c/d). Thus, the beat period
between these two periods should be around 2.2 days. Looking at the
{\sc theta} diagram of Kato et al. (2003a) shown in their Fig. 3, it is
easy to find the peak at frequency of 15.7 c/d, which can be associated
with the orbital period of the binary. The prewhitening of light curves of
Kato et al. (2003a) should certainly show that the peak at frequency of
15.7 c/d is real, corresponds to the orbital period and is responsible
for phase reversal.

\section{The August 2003 superoutburst}

Our observations from Aug 4 to Aug 8 caught the star at the end of the
superoutburst. The clear superhumps are visible in each night, including
the night of Aug 8 when the star faded almost to quiescence.

Fig. 11 shows the {\sc anova} power spectrum for the four nights of
the August superoutburst, with the overall brightness decrease removed.
The highest peak is found at the frequency $f=14.98\pm0.04$ c/d which
corresponds to the period of $P_{sh}=0.06674(18)$ days. This value
agrees well with the value of $P_{sh}$ obtained for the September
superoutburst. This is another argument for a constant superhump period
during the entire superoutburst, because in September we covered the
first and middle phases of the superoutburst and its late stages in
August.

\vspace{9.6cm}

\includegraphics{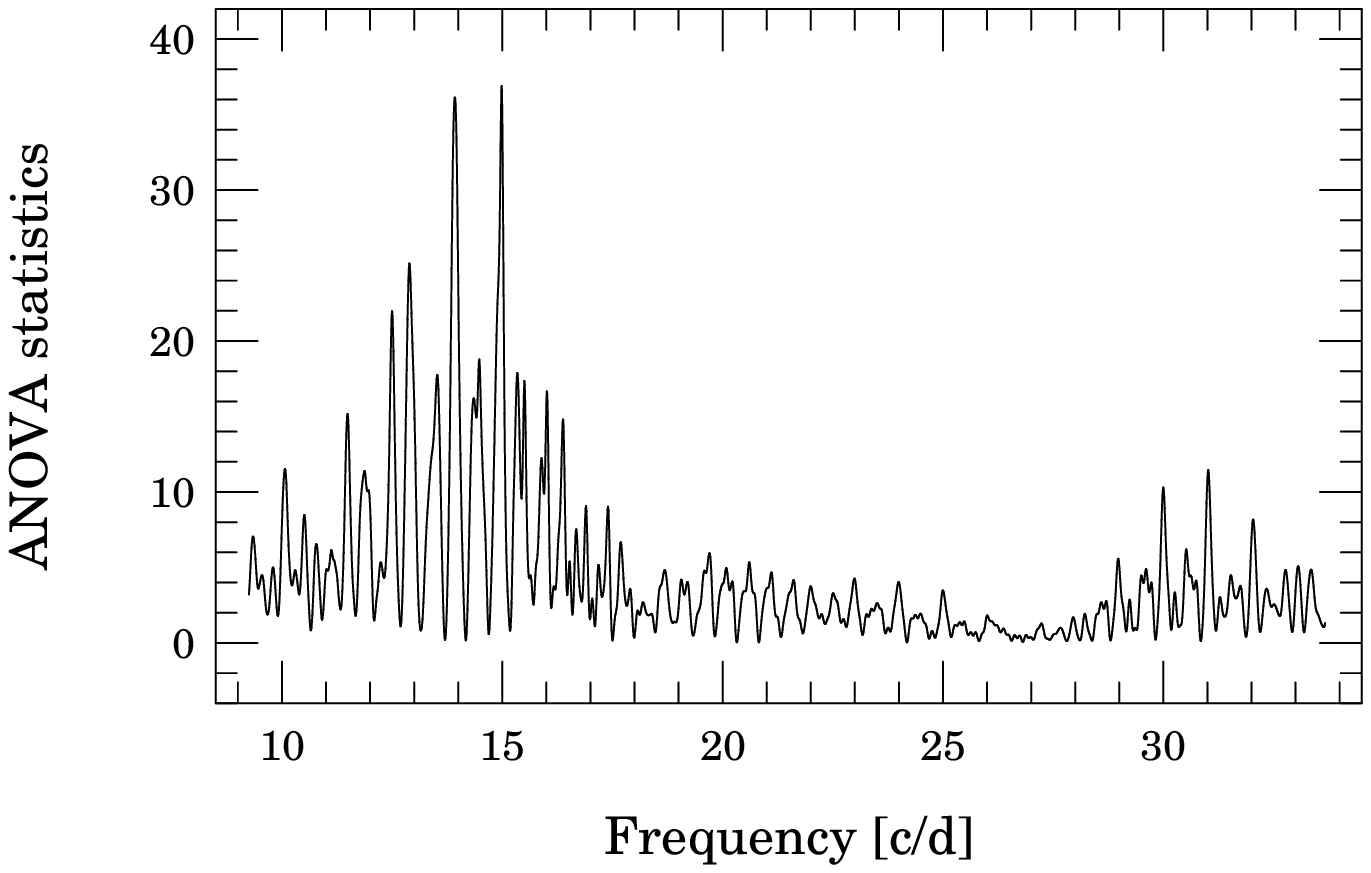}

   \begin{figure}[h]
      \caption{\sf {\sc Anova} power spectrum of the light curve of IX
Dra during its 2003 August superoutburst.
              }
   \end{figure}

\section{Nature of the second modulation and its implications}

\subsection{Evolution of CVs}

Typical dwarf nova starts its evolution as a binary stystem containing 
$\sim0.6 {\cal M}_\odot$ white dwarf and $0.2\div0.5 {\cal M}_\odot$
main sequence secondary. At the beginning, the orbital period is of
$\sim10$ hours and decreases due to angular momentum loss ($\dot J$)
through the magnetic braking via a magnetically constrained stellar wind
from the donor star (Hameury et al. 1988, Howell et al. 1997, Kolb and 
Baraffe 1999, Barker and Kolb 2003). At this stage the mass-transfer
rates are typically at the level of  $10^{-10}\div 10^{-8}~
\cal{M}_\odot ~{\rm yr}$$^{-1}$. The CV evolves towards shorter periods
until the secondary becomes completely convective and magnetic braking
greatly reduces. This happens for an orbital period of around 3 hours.
The mass transfer significantly decreases and the secondary shrinks
towards its equilibrium radius, below the Roche lobe. Abrupt
termination of magnetic braking at $P_{\rm orb}\sim 3$ hr produces a
sharp 2-3 hr period gap in dwarf novae distribution. The binary
reawakens as a CV at $P_{\rm orb}\sim 2$ hr when mass transfer
recommences. It is then driven mainly by angular momentum loss due to
the gravitational radiation ($\dot J_{\rm GR}$). The orbital period
still decreases while mass transfer stays at almost constant level.

Paczy\'nski (1981) was the first to find that the minimal period
for a CV containing a hydrogen rich secondary is around 80 min. 
Modern evolution codes locate this boundary at around 65 min. At this
point, the secondary star having mass of $0.06 {\cal M}_\odot$ becomes a
degenerate brown dwarf-like object and the system starts to evolve
towards longer orbital periods approaching $\sim 2$ hr within the Hubble
time. During this period the mass transfer is reduced to $\sim 10^{-12}~
{\cal M}_\odot ~{\rm yr}^{-1}$ and CV becomes a weak ($M_V \geq +12$
mag) and inactive star.

Evolution above the period gap is very fast, implying that most CVs are
presently at very short periods. Kolb (1993) predicted that 99\% of CVs
should be below the gap and 70\% of them should have "bounced" off the
minimum-period limit and now be evolving back toward longer periods.

This theoretical scenario seriously disagrees with observations,
especially with respect to the minimal period. A sharp period cut-off
is observed around 76 min. Currently, the three dwarf novae with shortest
orbital periods are GW Lib with 76.8 min, DI UMa with 78.6 min and V844
Her with 78.7 min (Fried et al. 1999, Thorstensen et al. 2002a).  Thus
theoretical models, placing period "bounce" at $\sim 65$ min, disagree
with observations at the level of $\sim10$\%. We do not deal here with
two clear outliers with extremely short periods: V485 Cen (Olech 1997)
and 1RXS J232953.9+062814 (Uemura et al. 2002, Thorstensen et al. 2002b)
which are most probably CVs which were formed with fairly old and
massive brown dwarf donors and thus can populate the period regime
shortwards of the "bounce" period (Kolb and Baraffe 1999).

There have been several hypothesis trying to explaining this
discrepancy. Patterson (1998, 2001) suggested that increasing the
angular momentum loss below the period gap to $\dot J \approx 2-3 \dot
J_{\rm GR}$ would solve the problem.  King et al. (2002) showed that an
intrinsic spread in minimum periods resulting from a genuine difference
in some parameter controlling the evolution can solve the problem. The
most probable second parameter might be different admixtures of magnetic
stellar wind braking in a small part of systems. Barker and Kolb (2003)
proposed that the additional source of angular momentum loss might be
the  mass loss from the system. However, the efficiency of this process
would have to be very large to move "bounce" period from 65 to 70 min.

Barker and Kolb (2003) also suggested that tidal deformation of the
secondary may have an effect on the period minimum. However, realistic
deformations increase the period minimum from 65 only to around 69 min.

It has also been suggested that the currently observed period minimum is
not a true minimum due to an age effect (King and Chenker 2002, Barker
and Kolb 2003). Simply, the systems which are currently at 75-77 min have
not had sufficient time to evolve to the true period "bounce".

The latest results of Andronov et al. (2003), based on data from open
clusters, show that empirical angular momentum loss from the secondary
is much longer than predicted by earlier models, and that the secondary
star is in thermal equilibrium above and in the period gap. Including
angular momentum loss from the secondary in the form suggested by
Andronov et al. (2003) improves the agreement between theory and
observations in determining the period minimum. Their data show that
angular momentum loss for systems below period gap is at the level of
1.5 larger than resulting from gravitational radiation alone. This is
still smaller that the value of 2-3 times suggested by Patterson (1998,
2001).

Another serious discrepancy between theory and observations lies in the
orbital period distribution. According to the theoretical models (Kolb
1993, Howell et al. 1997, Barker and Kolb 2003), the Galaxy is old
enough to have 99\% of its CVs below the period gap. As many as 70\% of them
should have reached their period minimum and to be currently evolving
towards longer periods. These CVs are predicted to have very low mass
transfer rates ($\dot M\leq 10^{-11}~{\cal M}_\odot ~ {\rm yr}^{-1}$)
and low time averaged absolute brightness ($M_V\geq +10$ mag). This
should create a clear period spike close to the minimum period, but
observations show that the period distribution below period gap is flat
(Patterson 1998).

In fact, we currently know only four CVs which can be assumed to be
potential period bouncers. They are WZ Sge, AL Com, EG Cnc and DI UMa.
The three former objects form a quite homogeneous group showing large
eruptions typically once per decade and no ordinary outbursts. Their
mass transfer rates are low and they are very faint objects with $M_V
\sim +12$ mag. On the other hand, DI UMa belongs to the very active
group of ER UMa stars. It goes into superoutburst every 30-45 days and
into ordinary outburst every 8 days (Kato et al. 1996, Fried et al.
1999) indicating very high mass transfer rate.

All these four objects have orbital periods very close to the period
cut-off. Thus, from the observational point of view there is no clear
evidence for the period "bounce". One can simply assume that there is a
larger dispersion in period excess for stars having short orbital
periods. A discovery of a star with the period excess below 1\% and an
orbital period significantly longer than 80 min would be strong evidence
for period "bounce".

\subsection{The period excess versus the orbital period}

The superhump period is simply the beat period between the orbital and
precession rate periods:

\begin{equation}
\frac{1}{P_{sh}} = \frac{1}{P_{orb}} - \frac{1}{P_{prec}}
\end{equation}

The precession rate of the eccentric disc was first discussed by Osaki
(1985). Based on a nonresonant free-particle orbit at the disk edge he
derived the following expression for the precession rate:

\begin{equation}
\frac{P_{orb}}{P_{prec}} =
\frac{3}{4}\frac{q}{\sqrt{1+q}}\left(\frac{R}{a}\right)^{3/2}
\end{equation}

\noindent where $a$ is the binary separation, $R$ is the disc radius and
$q$ is the mass ratio ${M_2}/{M_1}$. At the 3:1 resonance we can assume
that $R\approx0.46a$ and hence:

\begin{equation}
\frac{P_{orb}}{P_{prec}} \approx \frac{0.233q}{\sqrt{1+q}}
\end{equation}

Defining the period excess $\epsilon$ as:

\begin{equation}
\epsilon = \frac{\Delta P}{P_{orb}} = \frac{P_{sh} - P_{orb}}{P_{orb}}
\end{equation}

\noindent and using equation (5) we can simply derive the relation
between the period excess and the mass ratio:

\begin{equation}
\epsilon\approx\frac{0.23q}{1+0.27q}
\end{equation}

From the observational point of view it was first noticed by Stolz \&
Schoembs (1984) that $\epsilon$ grows with $P_{orb}$. This relation is
obeyed not only by the ordinary SU UMa stars but also by the permanent
superhumpers (Skillman \& Patterson 1993).

Direct measurements of mass ratio $q$ in CVs yields the following
relation:

\begin{equation}
\epsilon = 0.216(\pm0.018)q
\end{equation}

\noindent which roughly agrees with the theory (Patterson 2001).

Due to the mass loss from the secondary, the mass ratio $q$ decreases
with time. Thus it is very useful to trace the evolution of cataclysmic
variable stars in the $\epsilon - \log P_{orb}$ plane.

\subsection{IX Dra and its place in the family}

The longer period detected during the September superoutburst of IX Dra is
easily interpreted as the period of the superhumps. The most reasonable
explanation for the shorter period is that it is the orbital period of
the system. This then makes IX Dra a unique object for two reasons:

\begin{enumerate}
\item it is the first star, whose orbital period is visible throughout the
superoutburst, 
\item it has an unusually small period excess of $0.76\% \pm 0.03$\%
(a lower value has been observed only in EG Cnc - Patterson et al. 1998).
\end{enumerate}

Since the period excess is known to scale with mass ratio (see equations
7 and 8) the period difference observed in IX Dra corresponds to a mass
ratio of $q = 0.035\pm0.003$. Assuming a typical white dwarf mass of
between $0.6 - 0.8 {\cal M}_\odot$, the secondary component must have a
mass lower than $0.03 {\cal M}_\odot$. Even in the case of an extremely
massive white dwarf close to the Chandrasekhar limit, the secondary
component of IX Dra could not have a mass greater than 0.045 solar
masses, which makes it the best candidate among dwarf novae (alongside
EG Cnc) for a brown dwarf.

The only class of stars in which one observes changes with a period
equal to the orbital period during a superoutburst are WZ Sge stars, a
subgroup of SU UMa stars which have very rare outbursts and very low
mass transfer rates. In these stars, one does not observe superhumps
during the first $\sim 10$ days of a superoutburst, only changes with
orbital periodicity. Afterwards, the orbital humps are replaced by the
usual superhumps. It has recently been suggested that this is caused by
the low mass ratio in WZ Sge stars, which results in a very large disc
extending to the region where its particles can enter into a 2:1 orbital
resonance. This generates a two-armed spiral pattern of tidal
dissipation and is responsible for the light modulations with orbital
period. Superhumps are not seen from the beginning of the superoutburst
as the disc is not yet sufficiently elliptical because of the low rate of
mass transfer at this time (Osaki and Meyer 2002).

The unusually low mass ratio of IX Dra suggests that, during a
superoutburst, the accretion disc extends to 80\% of the distance
between the components of the system (see Figs. 1 and 2 in Osaki and
Meyer). This is analogous to the behavior of WZ Sge stars. However,
unlike in their case, IX Dra is characterized by a mass transfer rate at
least an order of magnitude higher. Thus, even at the start of the
superoutburst the disc is large, massive and eccentric, allowing both
the 2:1 and 3:1 resonances to occur. The disc precesses, and therefore
both orbital humps and superhumps can be observed throughout the entire
superoutburst.

Fig. 12 shows the position of IX Dra in the period excess vs. orbital
period diagram. Points mark normal SU UMa stars, squares correspond to
the permanent superhumpers while open circles indicate the WZ Sge dwarf
novae with recurrence times of $10-30$ years (WZ Sge, AL Com, EG Cnc)
and the star DI UMa, which has frequent outbursts like IX Dra. Two other
frequently outbursting dwarf novae: ER UMa and V1159 Ori are shown with
triangles. This figure is shown after Patterson (1998, 2001) with
over 20 new objects being added from Patterson et al. (2003) and Olech et
al. (2003b). The solid line shows the evolutionary path of a dwarf nova
with a white dwarf of mass $0.75 {\cal M}_\odot$ and a secondary
component with effective radius 6\% larger than that of a single
main sequence star (due to the distortion of the star filling its
Roche lobe - Renvoiz\'e et al. 2002).

\vspace{10cm}

\includegraphics{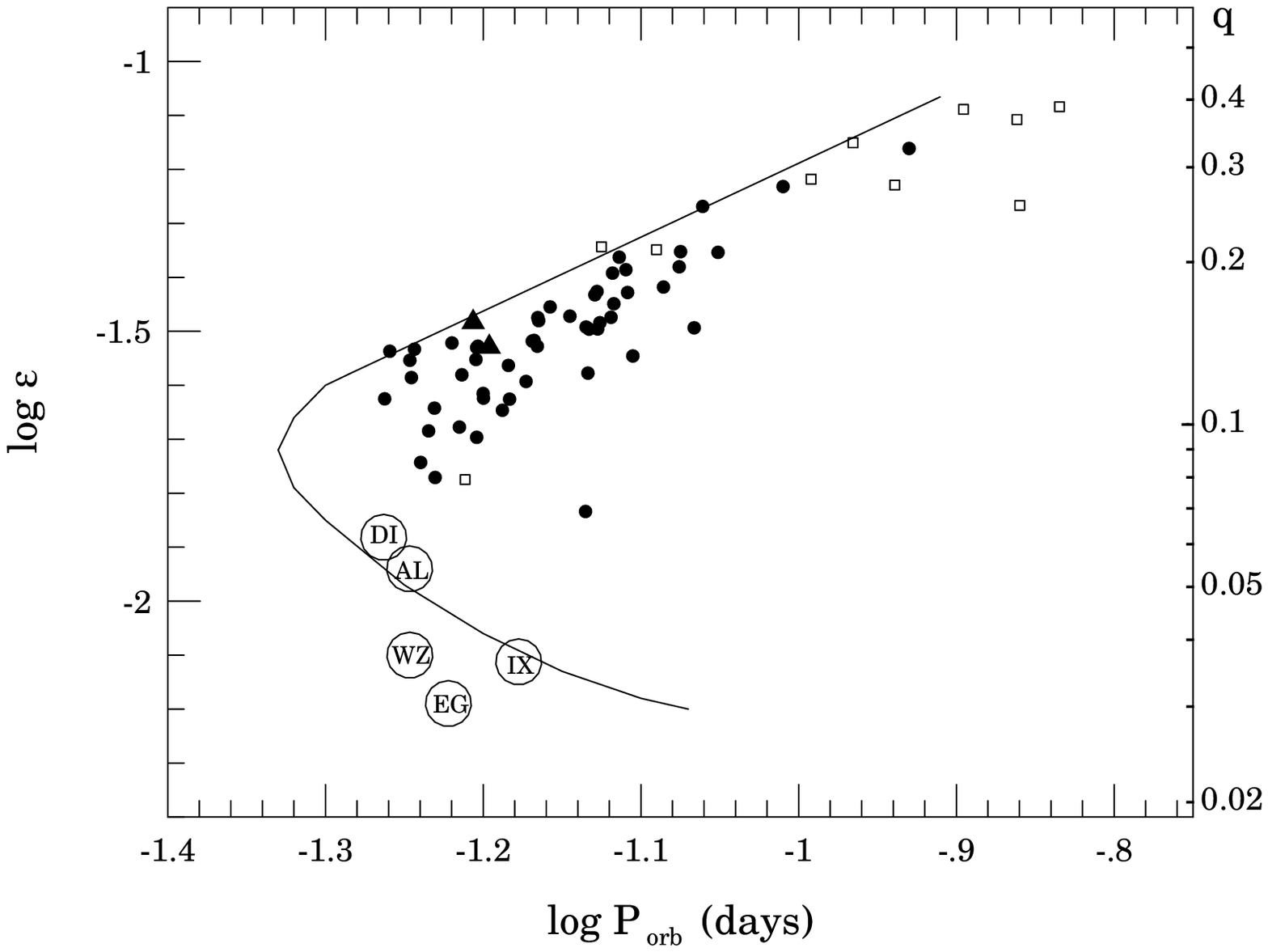}

   \begin{figure}[h]
      \caption {\sf The relation between the period excess and
orbital period of the system. The solid line corresponds to the
evolutionary track of a binary with a white dwarf of $0.75 {\cal M}_\odot$
and a secondary with effective radius 6\% larger than in the case of an 
ordinary main sequence star. Calculations are made under the
assumption that below the orbital period of two hours the angular
momentum loss in only due to gravitational radiation.}
\end{figure}

Looking at this picture we can see that the theoretical models do not
agree too well with observations. There are three ways of lowering the
theoretical line to better fit the observational data:

\begin{enumerate}
\item increasing the white dwarf mass up to $1.1-1.2$ solar masses, 
\item increasing the effective radius of the
distorted secondary by an unrealistic value of 20\% (Renvoiz\'e et al. 2002),
\item assuming that stars with periods below 2 hours lose angular
momentum about $2-3$ times more effectively than in the case of
gravitational waves alone (Patterson 1998, 2001).
\end{enumerate}

Hypothesis (3) is the most tempting, since it accelerates the evolution
of the system, moving the line down and increasing the minimum period
from 65-70 minutes up to 75 minutes, which agrees very well with
observations (the ordinary dwarf nova with the shortest known orbital
period is GW Lib with $P_{orb} = 76.78$ minutes - Thorstensen et al.
2002a).

Our discovery places IX Dra among the most evolved objects. According to
the theory they should have very low absolute brightness ($+12$ mag) and
very low accretion rates. However, both IX Dra and another atypical star, 
DI UMa, have the highest accretion rates observed in SU UMa stars,
resulting in their high level of activity in the form of frequent
outbursts and superoutbursts (Osaki 1995, Hellier 2001, Buat-M\'enard
and Hameury 2002). Therefore, we conclude that very old dwarf novae,
which most of the time are quiescent and behave like WZ Sge stars (with
outbursts every 20 years or more) occasionally show greatly increased
activity with a high accretion rate. DI UMa and IX Dra are currently
in this state. This high accretion rate may cause the mass loss from the
system, supplementing the emission of gravitational waves as the cause
of angular momentum loss.

\section{Normal outbursts}

In total, we detected seven ordinary outbursts of IX Dra. Three of them
with the best coverage are shown in Fig. 13. As in other SU UMa stars,
near the maximum of normal outburst IX Dra shows no other large
amplitude oscillations except small flickering. However at around 0.5
mag below the maximum the star often starts to show clear modulations.

The periodogram for nights from Aug 22 to Aug 25 is inconclusive, showing
many peaks of similar power in the frequency range 12 -- 16 c/d.

During the next outburst, which occurred on Aug 30, IX Dra was caught at
maximum light and during this night we did not detect any periodic
modulations. Our 3.9-h run on Aug 31 found the star at magnitude 0.8 mag
fainter, showing clear modulations with amplitude of around 0.25 mag.
The power spectrum for this run shows a broad peak at frequency of 15
c/d. The distance between two most prominent maxima is 0.132 d,
indicating a period of about 0.066 days.

The most conclusive results were obtained for the outburst lasting from
Sep 13 to Sep 16. The {\sc anova} power spectrum for these nights 
shows a clear peak at frequency of $15.06\pm0.07$ c/d, corresponding to
the period of $0.0664(3)$ days. This value is in very good agreement 
with the value of the orbital period derived from the September
superoutburst. On the other hand, it is only $2\sigma$ from the value
of superhump period.

An interesting feature visible in Fig. 13 are the clear dips
in the light curve of IX Dra observed at HJD of 874.44, 877.5 and
896.53. They appear aperiodically and their nature is unknown.

\clearpage

~

\vspace{19.9cm}

\includegraphics{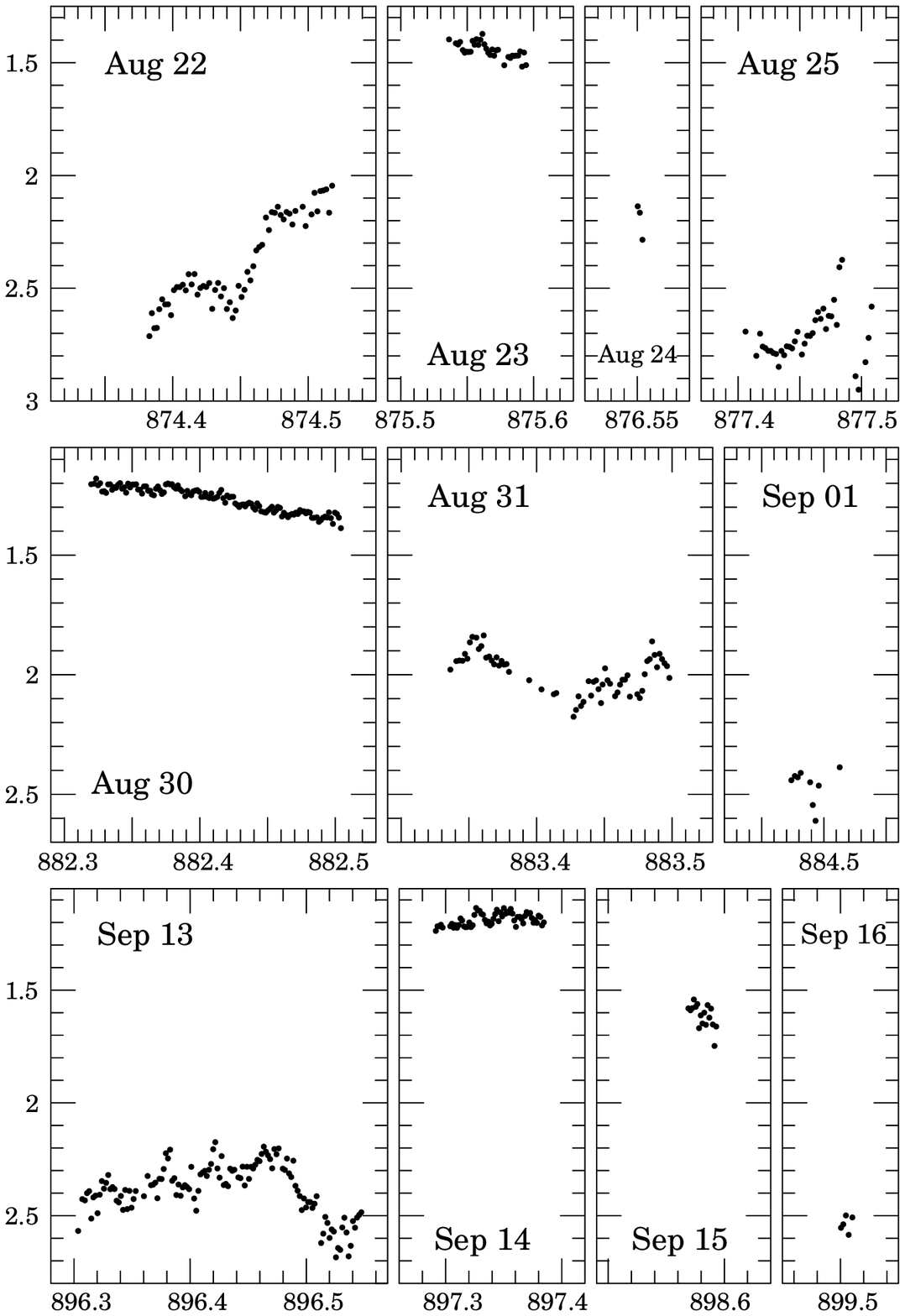}

   \begin{figure}[h]
      \caption {\sf The light curves of IX Dra during three normal
outbursts, whit the best observational coverage.}
\end{figure}

\clearpage

~

\vspace{20.5cm}

\includegraphics{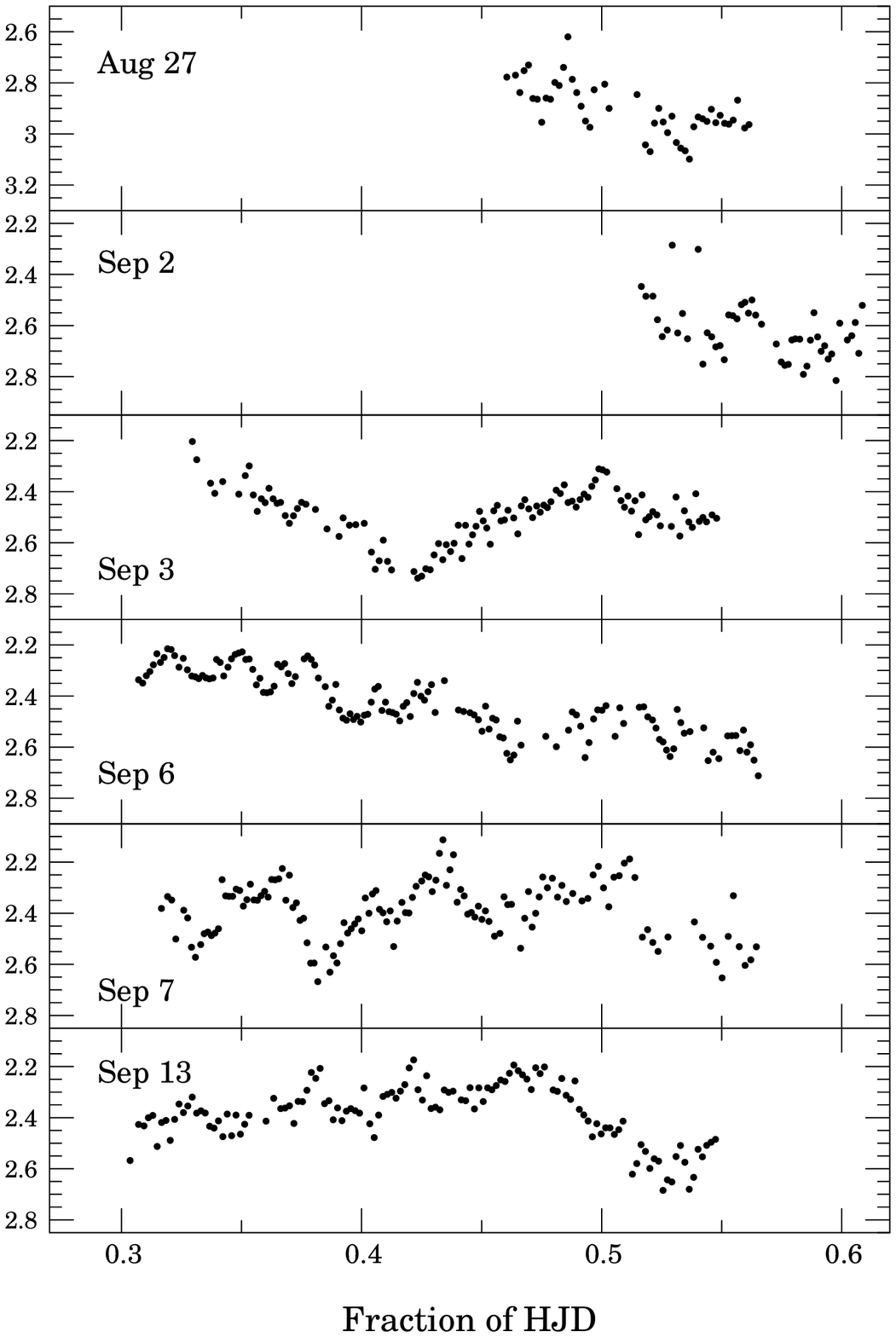}

   \begin{figure}[h]
      \caption {\sf The light curves of IX Dra from six longest runs
during quiescence.}
\end{figure}

\clearpage

Finally, we conclude that during ordinary outbursts IX Dra
occasionally shows orbital humps. However, we can not exclude that,
within errors, they are in fact superhumps or negative superhumps,
such as were detected in V1159 Ori during its outbursts (Patterson et
al. 1995).

\section{Quiescence}

Fig. 14 shows the light curves of IX Dra in quiescence obtained during
the six longest runs. The {\sc anova} power spectrum for 
interval Sep 6--13 is shown in the upper panel of Fig. 15. Before
calculation, the light curves were prewhitened using a first order
polynomial. It is clear that the highest peak corresponds to a frequency
of 13.05 c/d. However, we do not assume that this is the true period, but
rather a 2-day alias of the peak found at $f=15.03\pm0.03$. The latter
frequency corresponds to a period of $P_{orb}=0.0665(1)$. This value
agrees very well with our determination made for the September
superoutburst and in normal outbursts.

\vspace{11.5cm}

\includegraphics{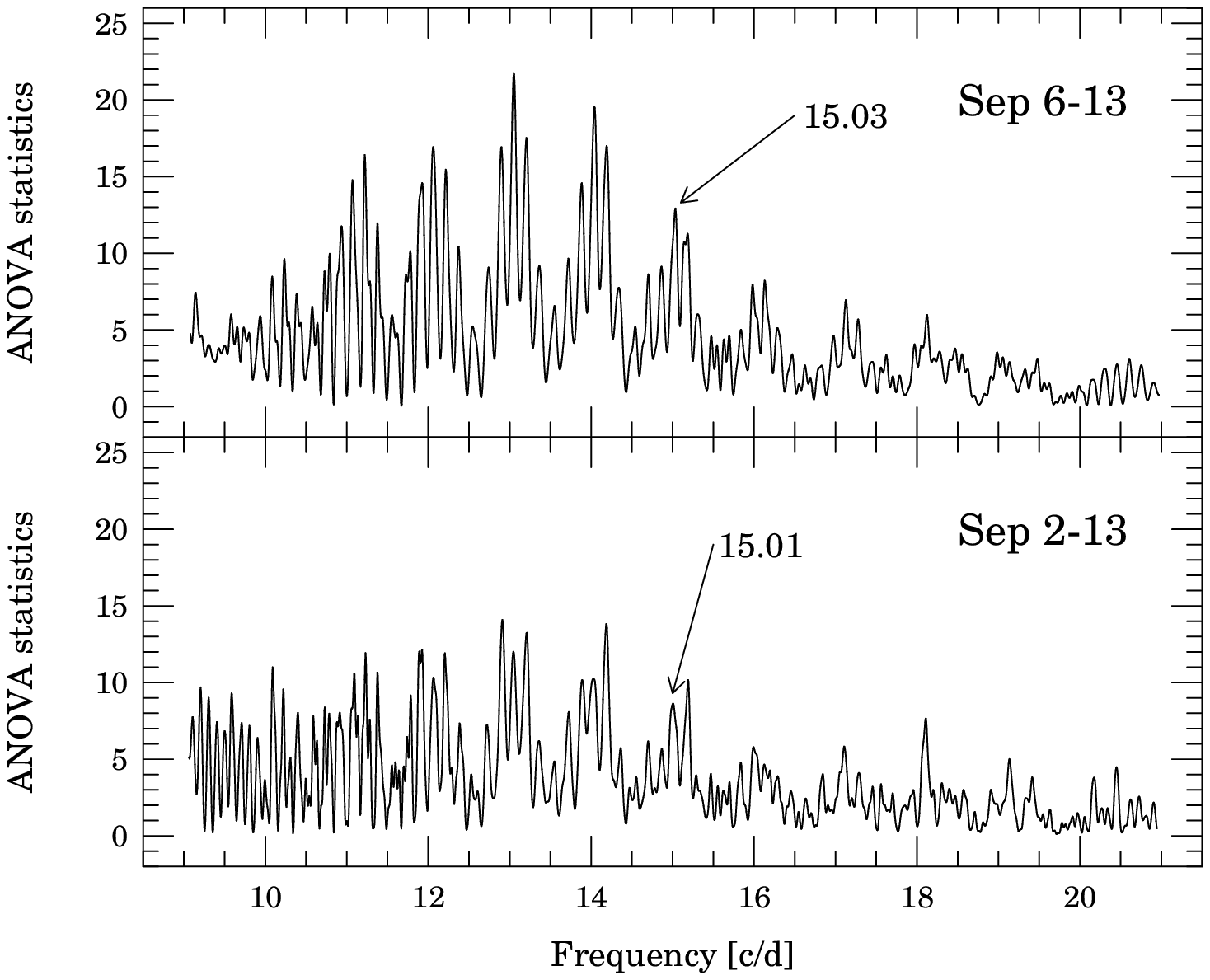}

   \begin{figure}[h]
      \caption {\sf The {\sc Anova} power spectrum for Sep 6--13
(upper panel) and Sep 2--13 (lower panel).}
\end{figure}

The question is why are the 1-day and 2-day aliases from Fig. 15 higher than
the true peak for both intervals Sep 6--13 and Sep 2--13. The
explanation might lie in high amplitude flickering, which masks the weak
orbital signal, and in mysterious wide dips which are clearly visible
during nights of Sep 3 and Sep 13. They add signal to the lower frequencies,
increasing low frequency aliases to a level higher than the main peak.

The main conclusion of this section is that, in quiescence, IX Dra
shows a weak signal with a period of $0.0665(1)$ days, i.e. with the same
value as was observed in the September 2003 superoutburst, supporting
the hypothesis that it is the orbital period of the binary system.

\section{IX Dra as a member of the ER UMa-type dwarf novae}

In the mid 1990s, when the first three members of the ER UMa group were
discovered, they seemed to be very unusual compared to normal SU
UMa stars. The ordinary SU UMa star with the shortest supercycle of 134
days was YZ Cnc (Patterson 1979, Shafter and Hessman 1988). Thus
supercycles of ER UMa stars were about 3-4 times shorter.
However, Patterson et al. (1995) describing the results of the CBA
observational campaign for V1159 Ori, claimed that there is no reason for
introducing a new class of variable stars. Simply, the observable traits
of ER UMa-type stars seem to be consistent with garden-variety SU UMa
stars. They follow the Kukarkin-Parengo relation connecting the amplitude
of the outburst with recurrence time between normal outbursts (Kukarkin
and Parenago 1934) and the Bailey relation connecting decay times from normal
eruptions and orbital period of the binary (Bailey 1975). They simply
appear to be normal SU UMa stars with greater activity and greater
luminosity due to their higher mass transfer rates (Osaki 1995).

However, careful inspection of the diagram with recurrence intervals for
supermaxima vs. normal maxima showed a significant gap between normal SU
UMa stars and ER UMa-type variables (see Fig. 18 of Paterson et al.
1995).

Since then, we have come to better understanding of  the ER UMa stars.
In particular, for four of them we know the orbital periods and we can
put them into the $\epsilon - P_{orb}$ plane. This is shown in Fig. 12,
where DI UMa and IX Dra are plotted with open circles and ER UMa and
V1159 Ori with triangles. It is clear that ER UMa stars occupy
completely different locations on this plane, having different mass
ratios and different evolutionary status.

\vspace{8.1cm}

\includegraphics{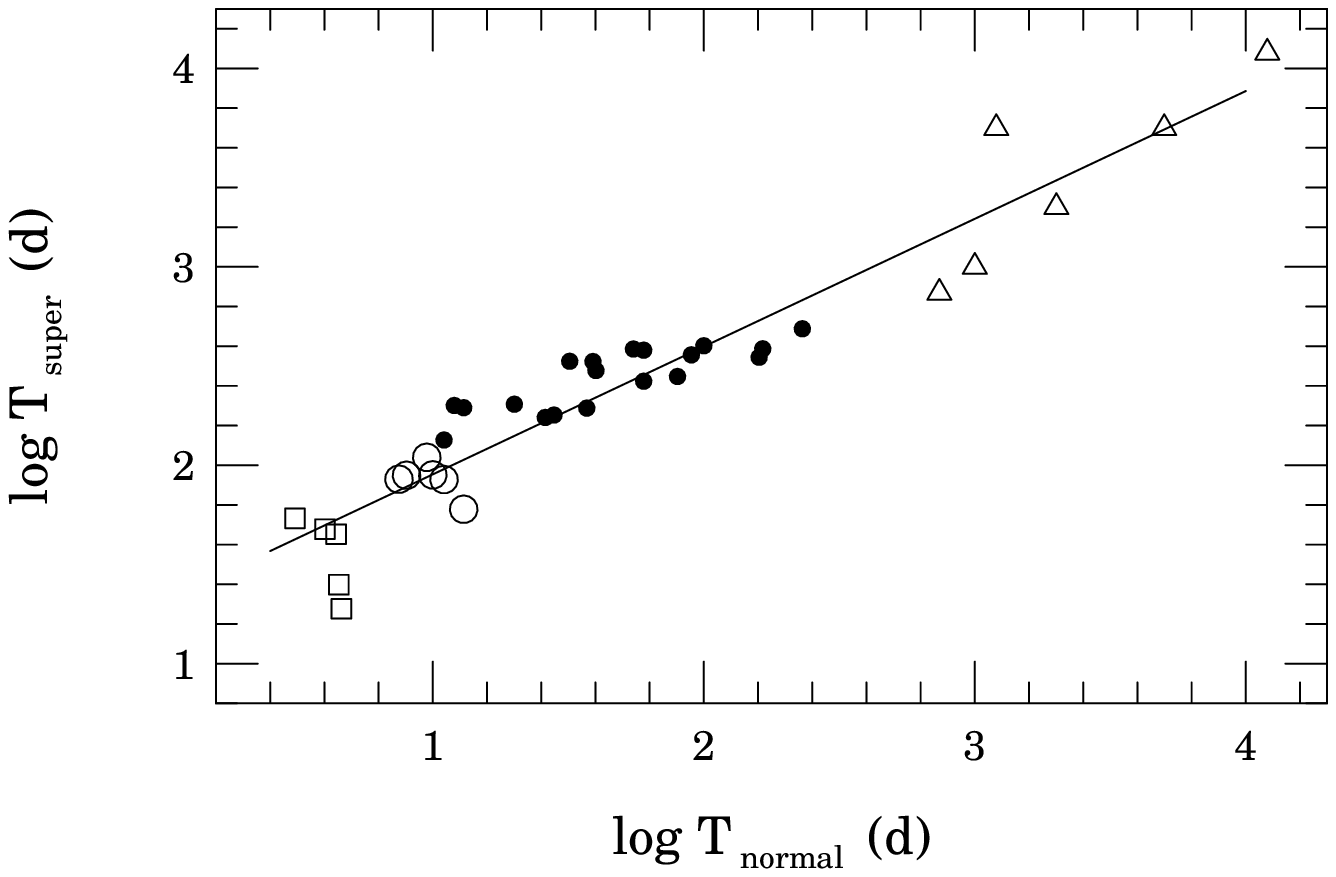}

   \begin{figure}[h]
      \caption {\sf The relation between recurrence interval between
superoutbursts and interval between normal outbursts. See text for
explanation of the symbols.}
\end{figure}

The discovery of new variables of the SU UMa-type fills the
above-mentioned gap in the diagram showing recurrence intervals for
supermaxima vs. normal maxima. Recent discoveries of MN Dra with a
supercycle of 60 days (Nogami et al. 2003), SS UMi and NY Ser with
supercycles of 85 days (Kato et al. 2000, Nogami et al. 1998), V503 Cyg
and BF Ara with supercycles of 89 days (Harvey et al. 1995, Kato et al.
2002a, 2003b) and V344 Lyr with supercycle of 109.6 days (Kato et al.
2002b) results in a smooth transition from the normal SU UMa stars to ER
UMa-type variables. This is shown in Fig. 16, where normal stars are
plotted with dots, WZ Sge variables with triangles, ER UMa stars with
squares and newly discovered normal SU UMa stars with circles. A simple
linear fit to the data gives the following relation:

\begin{equation}
\log T_s = 1.31 + 0.644 \log T_n
\end{equation}

\noindent which is shown as solid line. It is clear that this relation
describes very well both ER UMa and normal SU UMa stars.

Finally, the gap between normal SU UMa stars and ER UMa-type variables
has disappeared and, in the light of new discoveries, we can see a smooth
transition from one group to another. Our conclusion is similar to
the statement of Patterson et al (1995) that there is no need to artificially
distinguish ER UMa stars from other SU UMa variables.

\section{Summary}

We presented the results of the 2003 observational campaign of frequently
outbursting dwarf nova IX Draconis. The main conclusions of our work
are:

\begin{enumerate}

\item Analysis of the global light curve spanning five months of
observations shows that IX Dra is a very active object and goes into
superoutburst every 54 days and into a normal outburst every 3.1 days.
According to the model by Osaki (1995), this is due to a very high mass 
rate of mass transfer from the secondary.

\item During two best observed superoutburts, we detected the clear
superhumps with a period of \linebreak 0.066968(17) days ($96.43\pm0.02$
min). The superhump period seems to be constant during entire
superoutburst.

\item During September 2003 superoutburst, a second periodicity was
clearly visible in the light curve of the star. Its value was 0.06646(6)
days ($95.70\pm0.09$ min). A wave with the same period was detected
during normal outbursts and in quiescence. We suggest that this is the
orbital period of the binary system.

\item The beat between superhump and orbital period is the main cause of
an unusual phase reversal of maxima of superhumps - a phenomenon which
was previously observed also in ER UMa itself (Kato et al. 2003a). We
suggest that careful reanalysis of observations of ER UMa from
superoutburst obtained by Kato et al. (2003a) should result in discovery
of the orbital period.

\item The period difference observed in IX Dra corresponds to a mass
ratio $q = 0.035\pm0.003$. Assuming a typical white dwarf mass of
$0.6 - 0.8 {\cal M}_\odot$, the secondary component must have a
mass lower than $0.03 {\cal M}_\odot$, which makes it the best candidate
among dwarf novae (alongside EG Cnc) for a brown dwarf.

\item IX Draconis is the first SU UMa star showing orbital modulations
during the entire superoutburst. This is possibly due to the extremely
low mass ratio which allows the edge of the disc to reach 80\% of the
separation of the binary. In the outer region of the disc, its particles
can enter into a 2:1 orbital resonance. This generates a two-armed
spiral pattern of tidal dissipation and is responsible for the
appearance of the light modulations with orbital period (Osaki and Meyer
2002).

\item The position of IX Dra in the period excess vs. orbital period
plane suggests that the star is the most evolved cataclysmic variable
star in the Galaxy, which reached its period minimum long time ago and
now evolves towards longer periods.

\item High activity of IX Dra is in clear contrast with the behavior of
its neighbours in the $\epsilon - \log P_{orb}$ plane which are mostly
faint and inactive objects. Thus we suggest that very old dwarf novae,
which most of the time are quiescent and behave like WZ Sge stars, show
occasionally greatly increased activity with a high rate of accretion.
DI UMa and IX Dra are currently in this state. This high rate of
accretion may cause mass loss from the system, supplementing the
emission of gravitational waves as the cause of angular momentum loss.

\end{enumerate}

\bigskip \noindent {\bf Acknowledgments.} ~We acknowledge generous
allocation of  the Warsaw Observatory 0.6-m telescope time.  We would
like to thank Prof. J\'ozef Smak for reading and commenting on the
manuscript.

\end{document}